\documentclass[12pt]{amsart}
\usepackage{amsmath,amssymb, amscd, graphicx}

\usepackage{graphicx}
\usepackage{epstopdf}
\usepackage{amscd}
\usepackage{graphics}

\makeatletter
\@addtoreset{equation}{section}
\makeatother
\usepackage{enumerate}

\usepackage{color}

\def\CC{{\mathbb C}}

\def\RR{{\mathbb R}}
\def\ZZ{{\mathbb Z}}

\def\ee{{\mathrm e}}

\def\ii{{\mathfrak{i}}}

\def\cE{{\mathcal E}}
\def\cF{{\mathcal F}}

\def\cH{{\mathcal H}}

\def\cK{{\mathcal K}}

\def\cV{{\mathcal V}}

\def\cX{{\mathcal X}}
\def\cY{{\mathcal Y}}

\def\ft{{\mathfrak t}}

\def\tX{{\widetilde X}}

\def\tP{{\widetilde P}}

\def\tX{{\widetilde X}}

\def\tu{{\widetilde u}}

\def\tx{{\widetilde x}}
\def\ty{{\widetilde y}}
\def\tw{{\widetilde w}}

\def\tzeta{{\widetilde \zeta}}

\def\hy{{\widehat y}}

\def\res{\mathrm{res}}

\def\qed{\hbox{\vrule height6pt width3pt depth0pt}}

\def\nuI#1{{\nu^{\mathrm{I}}_{#1}}}

\def\nuII#1{{\nu^{\mathrm{II}}_{#1}}}

\def\trp{{\, {}^t\negthinspace}}

\def\book#1{\rm{#1}, }
\def\paper#1{\textit{#1}, }
\def\jour#1{\rm{#1}, }
\def\yr#1{({\rm{#1}) }}
\def\vol#1{\textbf{#1}}
\def\pages#1{\rm{#1}}
\def\page#1{\rm{#1}}

\def\publaddr#1{\rm{#1}, }
\def\publ#1{\rm{#1}, }
\def\by#1{{\rm{#1}, }}

\def\eds{\rm{eds.}}

\def\LF{\lfloor}
\def\RF{\rfloor}

\newtheorem{definition}{Definition}[section]

\newtheorem{theorem}[definition]{Theorem}
\newtheorem{proposition}[definition]{Proposition}
\newtheorem{corollary}[definition]{Corollary}
\newtheorem{remark}[definition]{Remark}
\newtheorem{lemma}[definition]{Lemma}

\def\book#1{\rm{#1}, }
\def\paper#1{\textit{#1}, }
\def\jour#1{\rm{#1}, }
\def\yr#1{({\rm{#1}) }}

\def\vol#1{\textbf{#1}}
\def\pages#1{\rm{#1}}
\def\page#1{\rm{#1}}

\def\publaddr#1{\rm{#1}, }
\def\publ#1{\rm{#1}, }
\def\by#1{{\rm{#1}, }}

\def\eds{\rm{eds.}}

\def\ft{u}

\begin{document}

\title{Nonlinear Schr\"odinger equation in terms of elliptic and hyperelliptic $\sigma$ functions}
\date{\today}

\author{Shigeki Matsutani}



\begin{abstract}
It is known that the elliptic function solutions of the nonlinear Schr\"odinger equation are reduced to the algebraic differential relation in terms of the Weierstrass sigma function,
$\displaystyle{
\left[-\frak{i}\frac{\partial}{\partial t}
+\alpha \frac{\partial}{\partial u}\right]\Psi 
-\frac{1}{2} \frac{\partial^2}{\partial u^2}\Psi +(\Psi^* \Psi) \Psi =
\frac12 (2\beta+\alpha^2-3\wp(v))\Psi
}$,
where $\Psi(u;v, t):=\mathrm{e}^{\alpha u+\ii\beta t+c}$ 
$\displaystyle{\frac{\mathrm{e}^{-\zeta(v)u}\sigma(u+v)}{\sigma(u)\sigma(v)}}$, its dual $\Psi^*(u; v,t)$, and certain complex numbers $\alpha, \beta$ and $c$.
In this paper, we generalize the algebraic differential relation to those of genera two and three in terms of the hyperelliptic sigma functions.
\end{abstract}

\maketitle
\bigskip

\tableofcontents
\section{Introduction}

Our purpose of this work is to investigate the algebraic solutions of the nonlinear Schr\"odinger (NLS) equation in terms of elliptic and hyperelliptic $\sigma$ functions.

As mentioned in Section 2, the elliptic solutions of the NLS equation are reduced to the algebraic differential relation \cite{K},
\begin{equation}
\left[-\ii\frac{\partial}{\partial t}
+\alpha \frac{\partial}{\partial u}\right]\Psi 
-\frac{1}{2} \frac{\partial^2}{\partial u^2}\Psi +(\Psi^* \Psi) \Psi =
\frac12 (2\beta+\alpha^2-3\wp(v))\Psi,
\label{eq:01}
\end{equation}
where 
$$
\Psi(u;v, t):=\ee^{\alpha u+\ii\beta t+c}\frac{\ee^{-\zeta(v)u}\sigma(u+v)}
{\sigma(u)\sigma(v)}, \quad 
\Psi^*(u):=-\ee^{-\alpha u-\ii\beta t-c}\frac{\ee^{-\zeta(v)u}\sigma(u-v)}
{\sigma(u)\sigma(v)}
$$
for the elliptic curve $E: \hy^2 = 4(x-e_1)(x-e_2)(x-e_3)$, $\alpha, \beta, c\in \CC$, and the Weierstrass sigma functions, $\sigma$, and zeta function, $\zeta$, of $E$.
After we fix the point $v$ and $e_i$ so that $\Psi^*$ is the complex conjugate of $\Psi$ and  $(\alpha, u)$ belongs to $\ii \RR\times \RR$ or $ \RR\times \ii\RR$, we find the elliptic solutions of the NLS equation.
We refer these conditions as the reality conditions of the NLS equation.
Determination of the conditions depends on the situations which we require, and is not simple because we tune the moduli parameter $e_i$ as in \cite{K}.

The construction of the algebraic solutions of the NLS equation was found by Previato 1985 \cite{Pr85} and is written in the book by Belokolos, Bobenko, Enolskii, Its and Matveev \cite{BBEIM} precisely.
On the other hand reevaluation of the theory of the hyperelliptic sigma function, which was discovered by Klein \cite{Klein86} and studied by Baker \cite{B1}, has been performed by several authors \cite{BEL, BEL20, EEL, Ma24S}.
We found the hyperelliptic solutions of the KdV equation \cite{BEL, Ma2}, MKdV equation \cite{Ma2},  Toda lattice equation \cite{KMP12}, sine-Gordon equation \cite{Mat05} in terms of the sigma functions and meromorphic functions of the symmetric product of algebraic curves.
However, hyperelliptic solutions of the NLS equation in terms of the hyperelliptic sigma functions have not yet been obtained.
When we obtain them, we can find the algebraic relations among the invariances of the solutions and numerically evaluate the time development for a certain class of curves as in \cite{Ma2, Ma23}; the evaluation is, sometimes, much easier than the approach by means of the theta functions.

In this paper, we leave aside the reality conditions for now  and present a generalization of the algebraic differential relation (\ref{eq:01}) of genera two and three, $g=2, 3$, in terms of the hyperelliptic sigma functions associated with the hyperelliptic solutions of the NLS equation.
To obtain this, we express the Baker function in terms of the sigma function reported in \cite{BEL}, though our Baker function differs from that in \cite{BEL}.
Using the Baker equation, we demonstrate this generalization.
We note that the Baker equation, or the generalized Lam\'e equation, of the Baker function contains the potential $\wp_{gg}$, while the addition theorem of the sigma functions gives the meromorphic functions of the Jacobian $J_X$, which does not coincide with $\wp_{gg}$ itself except for $g=1$.
Thus, this discrepancy prevents us from obtaining a generalization of (\ref{eq:01}).
Even for genera two and three, we find a generalization of the algebraic differential relation (\ref{eq:01}) in this paper.
In other words, the purpose of this paper is to show how to overcome the discrepancy so that we can generalize (\ref{eq:01}) associated with the hyperelliptic solution of the NLS equation.

We note that in the NLS equation, we have crucial problems related to the reality conditions depending on the {\lq\lq}nonfocusing" and {\lq\lq}focusing" NLS equations.
These correspond to the sign of the nonlinear terms in the NLS equation \cite{Pr85} and to the choice of $(\alpha, u)$ which belongs to $\ii \RR\times \RR$ or $ \RR\times \ii\RR$.
However, we will not address these problems at this stage, because we plan to consider them in the next step after considering the algebraic properties of the generalized algebraic differential relation.

The remainder of this paper is organized as follows.
Section 2 is devoted to the genus one case or the derivation of (\ref{eq:01}).
Section 3 presents a preliminary of the hyperelliptic sigma function.
Section 4 provides the addition formula related to the NLS equation.
In Section 5, we demonstrate the Baker functions in terms of hyperelliptic sigma functions.
In Section 6, we generalize (\ref{eq:01}) for the genera two and three cases using the Baker function and the addition formulae in Theorems \ref{thm:NLSg2} and \ref{thm:NLSg3}.

\section{Genus one case}\label{sec:2}
In this Section, we demonstrate how one gets an elliptic solution of the NLS equation (\ref{eq:01}) \cite{K}.
Let $E$ be  an elliptic curve given by
\begin{equation}
E:\
\frac{1}{4} \hy^2 = y^2 =x^3 + \lambda_2 x^2
                +\lambda_1 x + \lambda_0= (x-e_1)(x-e_2)(x-e_3),
\label{eq:ecurve}
\end{equation}
where the $e$'s are complex numbers, $\lambda_2 =-(e_1+e_2+e_3)$ and $\hy=2y$. 
In Weierstrass's convention, $\lambda_2=0$, but we leave it nonzero. 

The Weierstrass elliptic $\sigma$ function
associated with the curve $E$  is
connected with the Weierstrass $\wp$ and $\zeta$ functions
by
\begin{equation}
     \wp(u) =- \frac{d^2}{d u^2} \log \sigma(u), \quad
     \zeta(u) = \frac{d}{d u} \log \sigma(u).
\end{equation}
Here the coordinate $u$ in the Jacobian $J_E$ associated with $E$ 
is given by
\begin{equation}
     u = \int^{(x,y)}_\infty \frac{d x}{2 y},
\end{equation}
with $x(u)=\wp(u),~ \hy(u)=\wp'(u)$ and $\infty$
is the infinite point of $E$.
The Jacobian is given by $J_E = \CC/(2\ZZ \omega' + 2\ZZ \omega'')$
using the double period $(2\omega', 2\omega'')$ for
$\omega'=\displaystyle{\int_\infty^{(e_1,0)} \frac{d x}{2 y}}$ and 
$\omega''=\displaystyle{\int_\infty^{(e_3,0)} \frac{d x}{2 y}}$.

The key to obtain a $\wp$ function solution of the NLS equation is the addition formula,
\begin{equation}
        \wp(u) - \wp(v) = -
\frac{\sigma(u+v)\sigma(u-v)}{[\sigma(v)\sigma(u)]^2}.
\label{eq:add1}
\end{equation}
Let $(\wp(u), \wp'(u))=(x_1, 2y_1)$ and $(\wp(v), \wp'(v))=(x_2, 2y_2)$.
Using $\displaystyle{
\frac{d}{du}=2y_1 \frac{d}{d x_1}}$ and 
$\displaystyle{
\frac{d}{dv}=2y_2 \frac{d}{d x_2}}$, we find that  
(\ref{eq:add1}) leads to the following relations:
\begin{equation}
\zeta(u+v)+\zeta(u-v)-2 \zeta(u) = -\frac{2y_1}{x_2- x_1},
\label{3eq:g1zeta_uv1}
\end{equation}
\begin{equation}
\zeta(u+v)-\zeta(u)- \zeta(v) = \frac{y_2-y_1}{x_2- x_1},
\label{3eq:g1zeta_uv2}
\end{equation}
\begin{equation}
\wp(u+v)+\wp(u-v) = \frac{2 f_{\Omega}(x_1,x_2)}{(x_2- x_1)^2},
\label{3eq:g1wp_uv1}
\end{equation}
\begin{equation}
\wp(u-v) = \frac{f_{\Omega}(x_1,x_2)+2y_1y_2}{(x_2- x_1)^2},
\label{3eq:g1wp_uv2}
\end{equation}
\begin{equation}
\wp(u+v)+\wp(u)+\wp(v) = 
\left[\frac{y_2-y_1}{x_2- x_1}\right]^2-\lambda_2.
\label{3eq:g1wp_uv3}
\end{equation}
where 
\begin{equation}
f_{\Omega}(x_1,x_2)=x_1x_2(x_1+x_2+2\lambda_2) +\lambda_1(x_1+x_2)+\lambda_0.
\label{3eq:g1fOmega}
\end{equation}

By using (\ref{3eq:g1zeta_uv1})-(\ref{3eq:g1wp_uv3}) and by defining $\psi(u;v) = \displaystyle{\frac{\ee^{-\zeta(v) u} \sigma(u+v)}{\sigma(u)\sigma(v)}}$, we have the relation,
\begin{equation}
\left[\frac{d^2}{du^2}-2\wp(u)\right]\psi(u;v)=(\wp(v)+\lambda_2)\psi(u;v),
\label{4eq:BAeq_g1}
\end{equation}
which is known as the Lam\'e equation.
$\psi$ is known as the Baker function, or the Baker-Akhiezer function \cite{BEL,Ma24S}.

We also have $\displaystyle{\frac{d \psi(u;v)}{du}}=(\zeta(u+v)-\zeta(u)-\zeta(v)) \psi(u;v)$, or
\begin{equation}
\left[\frac{d}{du}-(\zeta(u+v)-\zeta(u))\right]\psi(u;v)=\zeta(v)\psi(u;v).
\label{4eq:BAeq_g1zata}
\end{equation}
Further, using the Miura relation, 
$
2x+\lambda_2+\tx = \mu^2 + \mu_u
$ for $\displaystyle{\mu:=\frac{d}{du}\log \psi(u;v) = \frac{y-\ty}{x-\tx}}$,
we obtain the static MKdV equation,
\begin{equation}
2(\lambda_2+3\tx)\mu_u+6\mu^2 \mu_u + \mu_{uuu}=0.
\label{3eq:MKdV_g2}
\end{equation}

We introduce
\begin{eqnarray*}
\Psi(u;v, t)&:=&\ee^{\alpha u+\ii\beta t+c}\psi(u;v)
=\frac{\ee^{\alpha u+\ii\beta t+c-\zeta(v) u} 
      \sigma(u+v)}{\sigma(u)\sigma(v)},\\
\Psi^*(u;v, t)&:=&\ee^{-\alpha u-\ii\beta t-c}\psi(u;-v)
=-\frac{\ee^{-\alpha u-\ii\beta t-c+\zeta(v) u} 
     \sigma(u-v)}{\sigma(u)\sigma(v)},
\end{eqnarray*}
for $\alpha, \beta, c\in \CC$.
Here we use $\sigma(-v)=-\sigma(v)$ and $\zeta(-v)=-\zeta(v)$.
We have 
\begin{equation*}
\left[-\ii\frac{\partial}{\partial t}
+\alpha \frac{\partial}{\partial u}\right]\Psi(u)
-\frac{1}{2} \frac{\partial^2}{\partial u^2}\Psi(u) +\wp(u) \Psi(u) =
\frac12 (2\beta+\alpha^2-\wp(v)-\lambda_2)\Psi(u).
\end{equation*}
Then the addition theorem (\ref{eq:add1}), $\Psi^* \Psi=\wp(u)- \wp(v)$, leads the relation,
\begin{equation}
\left[-\ii\frac{\partial}{\partial t}
+\alpha \frac{\partial}{\partial u}\right]\Psi 
-\frac{1}{2} \frac{\partial^2}{\partial u^2}\Psi +(\Psi^* \Psi) \Psi =
\frac12 (2\beta+\alpha^2-3\wp(v)-\lambda_2)\Psi,
\label{eq:NLSE1}
\end{equation}
which is the same as (\ref{eq:01}) when $\lambda_2=0$.


\section{Hyperelliptic curve $X$ and sigma functions}

In this Section, we provide a background information on the hyperelliptic $\theta$ functions and the $\sigma$ functions as a generalization of the Weierstrass elliptic $\sigma$ function.

\subsection{Geometrical setting of hyperelliptic curves}
Let $X$ be a hyperelliptic curve defined by
$$
	X~:~ y^2 = f(x):= x^{2g+1} + \lambda_{2g} x^{2g} + \cdots +\lambda_0
$$
together with a smooth point  $\infty$  at infinity.  
Here $\lambda$'s are complex numbers.
The affine ring related to $X$ is denoted as $R_X:=\CC[x, y]/(y^2 - f(x))$.
We fix the basis of the holomorphic one-form 
$$ 
    \nuI{i} =\frac{x^{i-1} d x}{2 y} \qquad (i=1,\ldots,g),
$$
and the homology basis for the curve  $X$,
$$
\mathrm{H}_1(X, \mathbb Z)
  =\bigoplus_{j=1}^g\mathbb Z\alpha_{j}
   \oplus\bigoplus_{j=1}^g\mathbb Z\beta_{j},
$$
so that their intersections are given by
$[\alpha_i, \alpha_j]=0$, $[\beta_i, \beta_j]=0$ and
$[\alpha_i, \beta_j]=-[\beta_i, \alpha_j]\delta_{ij}$.
We take the half-period matrix $\omega=\left[\begin{matrix}\omega'\\ \omega''\end{matrix}\right]$
 of  $X$  with respect to the given basis where
$$
    \omega'=\frac{1}{2}\left[\oint_{\alpha_{j}}\nuI{i}\right],
\quad
      \omega''=\frac{1}{2}\left[\oint_{\beta_{j}}\nuI{i}\right].
$$
The period matrices $2\omega'$  and  $2\omega''$ form the lattice $\Gamma_X$ in $\CC^g$ as a $\ZZ$-module.
The Jacobian variety of  $X$  is denoted by $J_X$, $\kappa_J: \CC^g \to J_X=\CC^g/\Gamma_X$.
The abelian covering of $X$ generated by the path space of $X$ with the basepoint $\infty\in X$ is denoted by $\tX$, $\kappa_X: \tX \to X$ $\kappa_X(\gamma_{P, \infty})=P$, where $\gamma_{P, \infty}$ is a path from $\infty$ to $P\in X$.
For a non-negative integer $k$, we define the Abelian integral $\tw$ from $k$-th symmetric product $S^k \tX$ of $\tX$ to $\CC^g$ and the Abel-Jacobi map $w$ from $S^k X$ by
\begin{equation*}
\tw:  S^k \tX \to \CC, \quad
  \tw(\gamma_1, \ldots, \gamma_k)= \sum_{i=1}^k
       \int_{\gamma_i} 
          \left[\begin{array}{c} \nuI{1} \\ \vdots 
                 \\ \nuI{g} \end{array}\right],
\end{equation*}
\begin{equation*}
w: S^k X \to J_X, \quad
  w((x_1,y_1), \cdots, (x_k, y_k))= \sum_{i=1}^k
       \int_\infty^{(x_i,y_i)} 
          \left[\begin{array}{c} \nuI{1}\\ \vdots \\ \nuI{g}
            \end{array}\right]
    \hbox{\rm mod}\ {\Lambda}. 
\end{equation*}
The image of $w$ is denoted by $W_X^k =  w(S^k X)$, i.e., $J_X=W_X^g$.
Furthermore we introduce an injection $\iota_X : X \to \tX$ and will fix it.
We find that $w = \kappa_J \circ \tw \circ \iota_X$.

\begin{definition}\label{4df:KdV_def2}
Let $(x_i, y_i)_{i=1, \ldots, g} \in S^g X$.
\begin{enumerate}
\item 
We define the polynomials associated with $F(x)=(x-x_1) \cdots (x-x_g)$ by
\begin{equation}
\pi_i(x) := \frac{F(x)}{x-x_i}=\chi_{i,g-1}x^{g-1} +\chi_{i,g-2} x^{g-2}
            +\cdots+\chi_{i,1}+\chi_{i,0},
\label{4eq:KdV_def2.1}
\end{equation}

\item We define $g\times g$ matrices as follows.
$$
 \cX_g := 
{\small{
\left(
\begin{array}{cccc}
     \chi_{1,0} & \chi_{1,1} & \cdots & \chi_{1,g-1}  \\
      \chi_{2,0} & \chi_{2,1} & \cdots & \chi_{2,g-1}  \\
   \vdots & \vdots & \ddots & \vdots  \\
    \chi_{g,0} & \chi_{g,1} & \cdots & \chi_{g,g-1}
     \end{array}\right)
}},\quad
\cY_g := 
{\small{
\left(\begin{array}{cccc}
     y_1 & \ & \ & \  \\
      \ & y_2& \ & \   \\
      \ & \ & \ddots   & \   \\
      \ & \ & \ & y_g  \end{array}
\right)}}=\frac{1}{2}\cH_g,
$$
$$
	\cF_g' := 
{\small{\left(
\begin{array}{cccc}F'(x_1)& &  &   \\
       & F'(x_2)&  &    \\
       &  &\ddots&    \\
       &  &  &F'(x_{g})\end{array}
\right)}},
\cV_g:= 
{\small{
\left(
\begin{array}{cccc} 1 & 1 & \cdots & 1 \\
                   x_1 & x_2 & \cdots & x_g \\
                   x_1^2 & x_2^2 & \cdots & x_g^2 \\
                    \cdot& \cdot &       & \cdot \\
                   x_1^{g-1} & x_2^{g-1} & \cdots & x_g^{g-1}
                 \end{array}\right)
}},
$$
$$
\cE_g:= 
{\small{
\left(
\begin{array}{ccccc}
      1         & \ & \ & \ &\  \\
      \varepsilon_{g-1} & 1 & \ &\mbox{\Large 0}  & \  \\
   \varepsilon_{g-2} &\varepsilon_{g-1}     & 1      &  & \ \\
    \vdots& \vdots          & \vdots & \ddots & \\
   \varepsilon_{1} &\varepsilon_{2}     & \cdots      & \varepsilon_{g-1} & 1 \\
     \end{array}\right)
}},
\quad
\cK_g := 
{\small{\left(
\begin{array}{cccc}
     x_1^{g-1} &   x_1^{g-2} & \cdots & 1  \\
  x_2^{g-1} &   x_2^{g-2} & \cdots & 1  \\
   \vdots & \vdots & \ddots & \vdots  \\
  x_g^{g-1} &   x_g^{g-2} & \cdots & 1
     \end{array}\right)
}},
$$
where $F(x)=x^g+\varepsilon_{g-1} x^{g-1}+\cdots \varepsilon_1 x + \varepsilon_0$ and $F'(x):=d F(x)/d x$.
\end{enumerate}

\end{definition}

\begin{lemma}\label{4lm:KdV1}
Let $u \in \tw(\iota_X((x_1, y_1), \ldots, (x_g, y_g)))$.

\begin{enumerate}

\item The inverse matrix of $\cX_g$ is given as $\cX_g^{-1}=\cF_g^{\prime-1} \cV_g$.

\item For $\partial_{u_i}:=\partial/\partial{u_i}$,
$\partial_{x_i}:=\partial/\partial{x_i}$, and
$\partial_{u_i}^{(r)}:=\partial/\partial{u_i^{(r)}}$, we have
$$
	{\small{
      \left[\begin{array}{c} \partial_{u_1}\\
                 \partial_{u_2}\\
                 \vdots\\
                 \partial_{u_g}
         \end{array}\right]
    }}
   =2 \cY_g\cF_g^{\prime -1}\cX_g
   {\small{
   \left[\begin{array}{c} \partial_{x_1}\\
                 \partial_{x_2}\\
                 \vdots\\
                 \partial_{x_g}
         \end{array}\right],}}
\quad
\frac{\partial x_i}{\partial u_r}=
\frac{2y_i}{F'(x_i)} \chi_{i, r-1}, \quad
\frac{\partial y_i}{\partial u_r}=
\frac{f'(x_i)}{F'(x_i)} \chi_{i,r-1},
$$
\begin{equation}
\frac{\partial}{\partial u_g }=
         \sum_{i=1}^g \frac{2y_i}{F'(x_i)} \frac{\partial}{\partial x_i},
           \quad
	\frac{\partial}{\partial u_{g-1} }=
         \sum_{i=1}^g \frac{2y_i\chi_{i,g-1}}{F'(x_i)}
              \frac{\partial}{\partial x_i}.
\label{4eq:hyp_dxdu}
\end{equation}
\end{enumerate}
\end{lemma}

\subsection{Sigma function and its derivatives}
We define differentials of the second kind,
$$
     \nuII{j}=\frac{1}{2y}\sum_{k=j}^{2g-j}(k+1-j)
      \lambda_{k+1+j} x^kdx,
     \quad (j=1, \cdots, g)
$$
and complete hyperelliptic integrals of the second kind
$$      \eta'=\frac{1}{2}\left[\oint_{\alpha_{j}}\nuII{i}\right], \quad
         \eta''=\frac{1}{2}\left[\oint_{\beta_{j}}\nuII{i}\right] .
$$
For this basis of the $2g$-dimensional space of meromorphic differentials, 
the half-periods $\omega',\omega'',\eta',\eta''$  satisfy   
the generalized Legendre relation
\begin{equation} 
{\mathfrak M}\left(\begin{array}{cc}0&-1_g\\1_g&0\end{array}\right) 
{\mathfrak M}^T= 
\frac{\ii\pi}{2}\left(\begin{array}{cc}0&-1_g\\1_g&0\end{array}\right),
\end{equation} 
where  $\mathfrak 
M=\left( 
\begin{array}{cc}\omega'&\omega''\\\eta'&\eta''\end{array} 
\right)$.
Let  $\tau={\omega'}^{-1}\omega''$.  
The theta function on  $\mathbb C^g$  with modulus  $\tau$ 
and characteristics $\tau a+b$ for $a,b\in \CC^g$ is given by
$$
    \theta\negthinspace\left[\begin{array}{c}a \\ b \end{array}\right]
     (z; \tau)
    =\sum_{n \in \mathbb Z^g} \exp \left[2\pi i\left\{
    \frac 12 \ ^t\,\negthinspace (n+a)\tau (n+a)
    + \ ^t\,\negthinspace (n+a)(z+b)\right\}\right].
$$ 
The $\sigma$-function (\cite{B1}, p.336,\cite{BEL}), 
an analytic function on the space  $\CC^g$  and a theta series having 
modular invariance 
of a given weight with respect to  $\mathfrak{M}$, 
is given by the formula
$$
 \sigma(u)
  =\gamma_0\, \mathrm{exp}\left\{-\frac{1}{2}\ ^t \ft
  \eta'{{\omega}'}^{-1}\ft   \right\}
  \theta\negthinspace
  \left[\begin{array}{c}\delta'' \\ \delta' \end{array}\right]
  \left(\frac{1}{2}{{\omega}'}^{-1}\ft \,;\, \mathbb T\right),
$$
where $\delta'$ and $\delta''$ are half-integer characteristics giving the vector of Riemann constants with basepoint at $\infty$   and $\gamma_0$ is a certain non-zero constant. 
The $\sigma$-function vanishes only on  $\kappa_J^{-1}(W_X^{g-1})$ 
(see for example \cite{B1}, p. 252). 
The hyperelliptic $\wp$ and $\zeta$ functions are defined by
$$
\wp_{i j} = -\frac{\partial^2}{\partial u_i \partial u_j} \log \sigma(u),
\quad
\zeta_{i} = \frac{\partial}{\partial u_i} \log \sigma(u).
$$
As the Jacobi inversion formula, we have the relation,
$$
F(x)=x^g -\sum_{i=1}^g \wp_{gi}(u) x^{i-1}.
$$

Let $\{\phi_i(x,y)\}$ be
an ordered  set of $\CC\cup\{\infty\}$-valued functions
 over $X$ defined by
\begin{equation}\label{eq:phi}
	\phi_i(x,y) = \left\{
        \begin{array} {llll}
     x^i                 \quad & \mbox{for } i \le g, \\
     x^{\LF(i-g)/2\RF+ g} \quad& \mbox{for } i > g~{\rm and}~ i - g\mbox{  even,}\\
     x^{\LF(i-g)/2\RF} y \quad & \mbox{for } i > g~{\rm  and}~ i - g\mbox{  odd.}\\
        \end{array} \right.
\end{equation}
Here we note that $\{\phi_i(x,y)\}$ is  a set of the bases of
$R_X$ as the $\CC$ vector space.

Following \cite{O},
we introduce a multi-index  $\natural^n$.  
For  $n$  with  $1\leq n<g$, we let  $\natural^n$  
be the set of positive integers  $i$  such that  
$n+1\leq i\leq g$  with  $i\equiv n+1$  mod  $2$.   
Namely,  
\begin{equation*}
\natural^n=\left\{\begin{array}{lll}
n+1, n+3, \cdots, g-1 \quad& \mbox{for}\; g-n\equiv 0\; \mathrm{mod}\; 2\\
n,   n+2, \cdots, g   \quad& \mbox{for}\; g-n\equiv 1\; \mathrm{mod}\; 2\\
 \end{array} \right.
\end{equation*}
and partial derivative over the multi-index $\natural^n$   
$$
	\sigma_{\natural^n}
        =\bigg(\prod_{i\in \natural^n}
	\frac{\partial}{\partial u_i}\bigg)  \sigma(u),
$$
For  $n\geq g$, we define  $\natural^n$  as empty and  $\sigma_{\natural^n}$  as  $\sigma$  itself.

\medskip
\centerline{Table 1}
\smallskip
  \centerline{
  \vbox{\offinterlineskip
    \baselineskip =5pt
    \tabskip = 1em
    \halign{&\hfil#\hfil \cr
     \noalign{\vskip 2pt}
     \noalign{\hrule height0.8pt}
& genus & \hfil\strut{\vrule depth 4pt}\hfil  & $ \sigma_{\natural^1}$ & $\sigma_{\natural^2}$ & $\sigma_{\natural^3}$  & $\sigma_{\natural^4}$ & $\sigma_{\natural^5}$ &$\sigma_{\natural^6}$&$\sigma_{\natural^7}$ &$\sigma_{\natural^8}$ &$\cdots$\cr
      \noalign{\hrule height0.3pt}
& $1$ & \strut\vrule  & $\sigma$        & $\sigma$       & $\sigma$       & $\sigma$      & $\sigma$      & $\sigma$   & $\sigma$   & $\sigma$ & $\cdots$ \cr
& $2$ & \strut\vrule  & $\sigma_2$      & $\sigma$       & $\sigma$       & $\sigma$      & $\sigma$      & $\sigma$   & $\sigma$   & $\sigma$ & $\cdots$ \cr
& $3$ & \strut\vrule  & $\sigma_2$      & $\sigma_3$     & $\sigma$       & $\sigma$      & $\sigma$      & $\sigma$   & $\sigma$   & $\sigma$ & $\cdots$ \cr
& $4$ & \strut\vrule  & $\sigma_{24}$   & $\sigma_3$     & $\sigma_4$     & $\sigma$      & $\sigma$      & $\sigma$   & $\sigma$   & $\sigma$ & $\cdots$ \cr
& $5$ & \strut\vrule  & $\sigma_{24}$   & $\sigma_{35}$  & $\sigma_4$     & $\sigma_5$    & $\sigma$      & $\sigma$   & $\sigma$   & $\sigma$ & $\cdots$ \cr
& $6$ & \strut\vrule  & $\sigma_{246}$  & $\sigma_{35}$  & $\sigma_{46}$  & $\sigma_5$    & $\sigma_6$    & $\sigma$   & $\sigma$   & $\sigma$ & $\cdots$ \cr
& $7$ & \strut\vrule  & $\sigma_{246}$  & $\sigma_{357}$ & $\sigma_{46}$  & $\sigma_{57}$ & $\sigma_6$    & $\sigma_7$ & $\sigma$   & $\sigma$ & $\cdots$ \cr
& $8$ & \strut\vrule  & $\sigma_{2468}$ & $\sigma_{357}$ & $\sigma_{468}$ & $\sigma_{57}$ & $\sigma_{68}$ & $\sigma_7$ & $\sigma_8$ & $\sigma$ & $\cdots$ \cr
& $\vdots$ & \hfil\strut {\vrule depth 5pt}\hfil  & $\vdots$ & $\vdots$ & $\vdots$ & $\vdots$ & $\vdots$  & $\vdots$   & $\vdots$   & $\vdots$ & $\ddots$ \cr
	\noalign{\hrule height0.8pt}
}}
}
\medskip

 For  $u\in\CC^g$, we denote by  $u'$  and  $u''$  
the unique vectors in  $\mathbb{R}^g$  such that 
  $$
  u=2\,{}^t\omega'u'+2\,{}^t\omega'' u''.
  $$
We define 
  \begin{eqnarray*}
  L(u,v)&=&{}^t{u}(2\,{}^t\eta'v'+2\,{}^t\eta''v''), \\ 
  \chi(\ell)&=&
  \exp\big\{2\pi i\big({}^t{\ell'}\delta''-{}^t{\ell''}\delta'
  +\frac12{}^t{\ell'}\ell''\big)\big\}
  \ (\in \{1, -1\})
  \end{eqnarray*}
for  $u$, $v\in\CC^g$  and for  $\ell$ 
($=2\,{}^t\omega'\ell'+2\,{}^t\omega''\ell''$) $\in\Lambda$.  
Then  $\sigma_{\natural^n}(u)$  for  $u\in\kappa_J^{-1}(W_X^n)$ 
 satisfies the translational relation
(\cite{O}, Lemma 7.3):
\begin{equation}
 \sigma_{\natural^n}(u+\ell)
   =\chi(\ell)\sigma_{\natural^n}(u)\exp L(u+\frac12\ell,\ell)\ \ 
 \hbox{for $u\in\kappa_J^{-1}(W_X^n)$}. 
\label{eq:trans}
\end{equation}
Then we also have the translational formulas of $\wp$ and $\zeta$ for $u\in \CC^g$,
\begin{equation}
\wp_{ij}(u+\ell)=\wp_{ij}(u),\quad
\zeta_{i}(u+\ell)=\zeta_{i}(u)
+2\sum_{j=1}^g(\eta_{ij}'\ell_j'+\eta_{ij}''\ell_j'').
\label{eq:wpzeta_g}
\end{equation}
Further for  $n\leqq g$,  we note that 
$\sigma_{\natural^n}(-u)=(-1)^{ng+\frac12n(n-1)}\sigma_{\natural^n}(u)$ 
for $u\in \kappa_J^{-1}(W_X^n)$, especially, 
\begin{equation}
\left\{
\begin{array}{llll}
\sigma_{\flat}(-u)=-\sigma_{\flat}(u) \ \ 
         & \hbox{for $u\in \kappa_J^{-1}(W_X^2)$} \\[1.5ex]
\sigma_{\sharp}(-u)=(-1)^g\sigma_{\sharp}(u) \ \ 
        &\hbox{for $u\in \kappa_J^{-1}(W_X^1)$}  
\end{array}\right.
\label{eq:sign}
\end{equation}
by Proposition 7.5 in \cite{O}.

\section{The addition formulae}
In this Section, we give the addition formulae of the hyperelliptic 
$\sigma$ functions which are the generalization
of the addition formula (\ref{eq:add1}). 
Those then provide the key to construct the hyperelliptic version of the algebraic differential relation (\ref{eq:01}) and (\ref{eq:NLSE1}).

Let us introduce the Frobenius-Stickelberger determinant:

\begin{definition}{\rm 
For a positive integer $n\geq 1$ and
 $(x_1,y_1), \cdots, (x_n, y_n)$ in $X$,
we define the Frobenius-Stickelberger determinant \cite{KMP12, Ma24S},
\begin{eqnarray*}
&&\Psi_n((x_1, y_1), \cdots, (x_n, y_n))\\
&& := 
\left|
\begin{array}{cccccc}
1 & \phi_1(x_{1}, y_{1}) &\cdots & \phi_{n-2}(x_{1}, y_{1}) & \phi_{n-1}(x_{1}, y_{1})  \\
1 & \phi_1(x_{2}, y_{2}) & \cdots & \phi_{n-2}(x_{2}, y_{2}) & \phi_{n-1}(x_{2}, y_{2})  \\
\vdots & \vdots & \ddots & \vdots & \vdots \\ 
1 & \phi_1(x_{n-1}, y_{n-1})  &\cdots & \phi_{n-2}(x_{n-1}, y_{n-1}) & \phi_{n-1}(x_{n-1}, y_{n-1})  \\
1 & \phi_1(x_{n}, y_{n})  & \cdots & \phi_{n-2}(x_{n}, y_{n}) & \phi_{n-1}(x_{n}, y_{n})  \\
\end{array}
\right|,\\
\end{eqnarray*}
where $\phi_i(x_j,y_j)$'s are the monomials defined in (\ref{eq:phi}).
}
\end{definition}

We have the addition formula for the hyperelliptic $\sigma$ functions as the generalization of (\ref{eq:add1}).
(Theorem 5.1 in \cite{EEMOP}):
\begin{theorem}\label{thm:add} 
Assume that $(m, n)$ is a pair of positive integers. 
Let  $(x_i,y_i)$  $(i=1, \cdots, m)$, $(x'_j,y'_j)$ $(j=1, \cdots, n)$  in  
$X$ 
and  $u\in\kappa_J^{-1}(W_X^m)$, $v\in\kappa_J^{-1}(W_X^n)$  be points 
such that  $u=\tw(\iota_X(x_1, y_1),\cdots,\iota_X(x_m,y_m))$ 
and  $v=\tw(\iota_X(x_1', y_1'),\cdots, \iota_X(x_n',y_n'))$.  
Then the following relation holds\,{\rm :}
\begin{eqnarray}
&&\frac{\sigma_{\natural^{m+n}}(u + v) \sigma_{\natural^{m+n}}(u - v) }
{\sigma_{\natural^m}(u)^2 \sigma_{\natural^n}(v)^2} \nonumber\\
&&=\delta(g,m,n)
\frac{\prod_{i=0}^1 \Psi_{m+n}
((x_1,y_1),\cdots, (x_m, y_m), 
(x_1',(-1)^i y_1'),\cdots,(x_n',(-1)^i y_n'))}
{(\Psi_{m}((x_1,y_1)\cdots, (x_m, y_m))
\Psi_{n}((x_1',y_1')\cdots, (x_n', y_n')))^2}\nonumber\\
&&\times
\prod_{i=1}^m\prod_{j=1}^n 
\frac{1}{ \Psi_2((x_i, y_i), (x_j', y'_j))},
\label{eq:Thadd}
\end{eqnarray}
where $\delta(g,m,n)=(-1)^{gn+\frac12n(n-1)+mn}$.
\end{theorem}

Theorem \ref{thm:add} with $m=g$ and $n=1$ leads to the following Corollary, which Belokolos, Enolskii, and Salerno found \cite[Theorem 3.2]{BES};
\begin{corollary} \label{cor:deg2}
Let  $(x_i,y_i) \in X$  $(i=1, \cdots, g)$, $(x, y):=(x'_1,y'_1)\in X$,
 $u\in \CC^g$ and $v\in\kappa_J^{-1}(W_X^1)$ such that  $u=w((x_1, y_1),\cdots,(x_g,y_g))$ modulo $\Gamma_X$ and  $v=w((x, y))$ modulo $\Gamma_X$.  
Then we have 
\begin{eqnarray}
\frac{\sigma(u + v) \sigma(u - v) }
{\sigma(u)^2 \sigma_{\natural^1}(v)^2} 
 &=&{x}^g - \sum_{i =1}^g \wp_{g j}(u)
 x^{i -1}\nonumber \\
 &=&F(x) = (x - x_1) (x - x_2) \cdots (x - x_g).
\label{eq:addthem_F}
\end{eqnarray}
\end{corollary}

\noindent
{\bf{Proof:\ }}
Direct computation shows it.
\qed

Considering the difference in the differential operators in Jacobian $J_X$,
we generalize (\ref{3eq:g1zeta_uv1})-(\ref{3eq:g1wp_uv3}).
Let  $(x_i,y_i) \in X$  $(i=1, \cdots, g)$ and $(x,y) \in X$ as follows.

\begin{lemma}\label{3lm:addhyp4}
For $u=\tw(\iota_X(x_1, y_1), \ldots, \iota_X(x_g,y_g))$ and $v=\tw(\iota_X(x, y))$ in the case of $k=g$ of Corollary \ref{cor:deg2}, the following hold.
\begin{enumerate}
\item 
$\displaystyle{
\sum_{i=1}^g e^{i-1}
\left(
\zeta_i(u+v) +\zeta_i(u-v) -2 \zeta_i(u)\right)
=- \sum_{\ell=1}^g\frac{2 y_\ell F(e)}{(e-x_\ell)(x-x_\ell)F'(x_\ell)},
}$
\item 
$\displaystyle{
\zeta_i(u+v) +\zeta_i(u-v) -2 \zeta_i(u)
=-\sum_{\ell=1}^g  \frac{2 y_\ell \chi_{\ell, i-1}}{(x-x_\ell)F'(x_\ell)},
}$
\item 
$\displaystyle{
\sum_{i=1}^g x_r^{i-1}
\left(
\zeta_i(u+v) +\zeta_i(u-v) -2 \zeta_i(u)\right)
=-\frac{2 y_r}{(x-x_r)},
}$
\item 
$\displaystyle{
\sum_{i=1}^g x^{i-1}
\left(
\zeta_i(u+v) -\zeta_i(u-v) -2 \zeta_i^{\natural_1}(v)
\right)
=\frac{2y F'(x)}{F(x)}
=\sum_{\ell=1}^g \frac{2y}{(x-x_\ell)},
}$
\item 
$\displaystyle{
\sum_{i,j}^g x^{i-1} x_r^{i-1}
\left(
\wp_{i j}(u+v) -\wp_{i j}(u-v) \right)
= 
-\frac{4y_r y }
{(x-x_r)^2}.
}$
\end{enumerate}
where $e$ is a generic parameter.
Further we let $\displaystyle{
\zeta_i^{\natural_j}(v):=\left[\frac{\partial}{\partial \tu_i} \log\sigma_{\natural^j}(\tu)\right]_{u=v}}$ for $\tu \in \CC^g$.
\end{lemma}

\noindent
{\bf{Proof:\ }}
Lemma \ref{4lm:KdV1} and (\ref{4eq:KdV_def2.1}) show 1.
Noting $\partial/\partial x = \sum_i(2y/x^{i-1}) \partial/\partial v_i$, we have {3}, {4}, and {5}.
\qed

\begin{proposition}\label{5pr:Baker_wp1}
Let $u=\tw(\iota_X(x_1, y_1), \ldots, \iota_X(x_g,y_g))$, and $v=\tw(\iota_X(x,y))$.
\begin{eqnarray}
\wp_{gg}(u+v)
&+&\wp_{gg}(u-v)+2\wp_{gg}(u) =
\nonumber \\
&2&\left[
\sum_{r} \frac{y_r}{(x-x_r)F'(x_r)}\right]^2
+2\frac{f(x)}{F(x)^2}-2\lambda_{2g}-2 x.
\label{3eq:g_wp_uv3}
\end{eqnarray}
\end{proposition}

\noindent
{\bf{Proof:\ }}
We consider the differential of Lemma \ref{3lm:addhyp4} (ii) with respect to $u$.
We interpret the differential (\ref{4lm:KdV1}) with respect to $u$ into differentials with respect to $x_r$, then we have
\begin{eqnarray*}
\wp_{gg}(u+v)&+&\wp_{gg}(u-v)-2\wp_{gg}(u) 
=\sum_{\ell} \frac{2y_\ell}{F'(x_\ell)}\frac{\partial}{\partial x_\ell}
\sum_r \frac{2y_r}{(x-x_r)F'(x_r)}\\
&=&2\left[
\sum_{r} \frac{y_r}{(x-x_r)F'(x_r)}\right]^2
+\sum_r\frac{2}{F'(x_r)}\frac{\partial}{\partial x}
\left[\frac{f(x)}{(x-x)F'(x)}\right]\Bigr|_{x=x_r} \nonumber \\
&=&2\left[\sum_{r} \frac{y_r}{(x-x_r)F'(x_r)}\right]^2
+2\left[\frac{f(x)}{F(x)^2}-\lambda_{2g}- x-2\wp_{gg}(u)\right].\\
\end{eqnarray*}
The second equality is asserted by Lemma \ref{4lm:d2F/dx2}.
Lemma \ref{4lm:MKdV01_part1} shows the final step.
\qed

\begin{lemma}\label{4lm:d2F/dx2}
$\displaystyle{
	\frac{\partial}{\partial x_k} F'(x_k)
         =\frac{1}{2}
 \left[\frac{\partial^2}{\partial x^2} F(x)\right]_{x=x_k}
}$.
\end{lemma}

\begin{lemma}\label{4lm:MKdV01_part1}
For $(\tx, \ty)\in X$ and $F(x)$, we have
$$
\displaystyle{
\sum_{k=1}^g \frac{1}{F'(x_k)}
             \left[\frac{\partial}{\partial x}\left(
       \frac{f(x)}{(x - \tx) F'(x)} \right) \right]_{x = x_k}
       =-\frac{f(\tx)}{F(\tx)^2}+ \lambda_{2g} + b_r + 2 \wp_{g g}. 
}
$$
\end{lemma}

\noindent
{\bf{Proof:\ }}
The relation is proved by the integral.
\begin{enumerate}

\item 
Let us consider the polygon expansion of the curve $X$ by $D_X$ associated with the standard homology basis.
Let its boundary be denoted by $\partial D_X$.
\begin{equation}
  \oint_{\partial D_X} \frac{f(x)}{(x-\tx)F(x)^2} d x =0.
\label{4eq:MKdV_0012}
\end{equation}

\item  The divisor of the integrand in (\ref{4eq:MKdV_0012}) is 
\begin{eqnarray}
\left(\frac{f(x)}{(x-b_r)F(x)^2} d x\right) &=&
    3 \sum_{i=1, b_i}^{2g+1} (b_i,0) -(\tx,\ty)-(\tx,-\ty)\nonumber\\
&-&
        2\sum_{i=1}^g (x_i,y_i) -
        2\sum_{i=1}^g (x_i,-y_i) - 3\infty .
\end{eqnarray}

\item By considering the local parameter $t$ ($x=1/t^2$) at $\infty$, we have
$$
\res_{\infty}\frac{f(x)}{(x-\tx)F(x)^2} d x
      = -2 (\lambda_{2g} + \tx + 2 \wp_{g g}).
$$

\item   By considering the local parameter $t$ ($t=x-\tx$) at $(\tx,\pm \ty)$, we obtain
$$
\res_{(\tx,\pm \ty)}\frac{f(x)}{(x-\tx)F(x)^2} d x
      =  \frac{f(\tx)}{F(\tx)^2}
        . 
$$

\item  By considering the local parameter $t$ ($t=x-x_k$) at $(x_k,\pm y_k)$, we have
$$
\res_{(x_k,\pm y_k)}\frac{f(x)}{(x-\tx)F(x)^2} d x
      =  \frac{1}{F'(x_k)}
             \left[\frac{\partial}{\partial x}\left(
       \frac{f(x)}{(x - \tx) F'(x) }\right) \right]_{x = x_k}
        . 
$$

\end{enumerate}
Using them we prove the relation.
\qed

Here we introduce $\displaystyle{\tzeta_i(P) := -\int^P_\infty \nuII{i}}$.

\begin{lemma}\label{4lm:zetazeta=mero}
Let $u=\tw(\iota_X(x_1, y_1), \ldots, \iota_X(x_g,y_g))$, $v=\tw(\iota_X(x,y))$.Then we have
\begin{equation}
[\zeta_g(u+v)-\zeta_g(u-v)-2\tzeta_g(v)]=\frac{2y}{F(x)}.
\label{4eq:zeta_Anstz}
\end{equation}
\end{lemma}

\noindent
{\bf{Proof:\ }}
The left hand side is the meromorphic function on $J_X\times W_X^1$ because of the translational formula (\ref{eq:wpzeta_g}).
The right hand side is a meromorphic function on $S^gX \times X$ and thus, on $J_X\times W_X^1$ by the Abel-Jacobi theorem. 
Thus we show the equality of both sides.
We compare the divisors.
As a function of $v$, both sides vanish at every finite branch point $(x,y)=(b_r,0)$ from the translational formula (\ref{eq:wpzeta_g}) and the definition of $\tzeta_g$.
At $v=0$ or $P=\infty$, both sides diverge in order of one.
At $v=u$ or $v=-u$, correspoing to $x=x_i$, the order of the pole is one.
We note that both sides are odd functions with respect to $v$.
For $v$, we expand both sides at $v=0$,
$$
\zeta_g(u+v)-\zeta_g(u-v) = -2\sum_{j} \wp_{gj} v_j 
-\frac{2}{3!}\sum_{jk} \wp_{gjk\ell} v_j v_k v_\ell
 + \mbox{the oder of }v \ge 3
$$
The local parameter $t = x^{g+1}/y$ shows 
$$
\zeta_g(u+v)-\zeta_g(u-v)-2\tzeta_g(v)
 = 2(x_1+\cdots+x_g) t +\frac{2}{t} +\mbox{the oder of }t >0.
$$
$$
\frac{2\ty}{F(\tx)}=\frac{2}{t}
\left(1+(x_1+\cdots +x_g)t^2+\mbox{the oder of }t >0\right).
$$ 
The leading terms of both sides match. Hence we prove it.
\qed

\begin{remark}{\rm{
Lemma \ref{4lm:zetazeta=mero} may be generalized as follows.
\begin{eqnarray}
&&[\zeta_{g-1}(u+v)-\zeta_{g-1}(u-v)+2\tzeta_{g-1}(v)]
=\frac{2y}{F(x)}(x-\wp_{gg}),\nonumber
\\
&&[\zeta_{g-2}(u+v)-\zeta_{g-2}(u-v)+2\tzeta_{g-2}(v)]
=\frac{2y}{F(x)}(x^2-\wp_{gg}x+\wp_{g,g-1}).
\end{eqnarray}
}}
\end{remark}

Using above relations, we have the following lemma:
\begin{lemma}\label{5lm:Baker_wp2}
For $u=\tw(\iota_X(x_1, y_1), \ldots, \iota_X(x_g,y_g))$ and $v=\tw(\iota_X(x,y))$
$$
\zeta_g(u+v) - \zeta_g(u)-\tzeta_{g}(v)
=\frac{y}{F(x)}
-\sum_{r=1}^g  \frac{ y_r}{(x-x_r)F'(x_r)},
$$
$$
\wp_{gg}(u+v) - \wp_{gg}(u-v)=-\sum_{r=1}^g
\frac{4y y_r}{(x-x_r)F'(x_r)F(x)},
$$
$$
\wp_{gg}(u+v)+\wp_{gg}(u)+x =
\left[
\sum_{r} \frac{y_r}{(x-x_r)F'(x_r)}-\frac{y}{F(x)}\right]^2
-\lambda_{2g}.
$$
\end{lemma}

\section{Baker function for $X$}

Let us generalize (\ref{4eq:BAeq_g1}) and 
$\psi(u;v)=\displaystyle{\frac{\ee^{-\zeta(v) u} \sigma(u+v)}{\sigma(u)\sigma(v)}}$ of genus one to general $g$.

Let us define
\begin{equation}
\psi(u;v)=\frac{\ee^{-\sum_i \tzeta_i(P) u_i} 
\sigma(u+v)}{\sigma(u)\sigma_{\natural_1}(v)}
\label{4eq:Bakereq}
\end{equation}
Here $\displaystyle{\tzeta_i(P):= -\int^P_\infty \nuII{i}}$ for $u\in \CC^g$ and $P \in X$, and $v=\tw(\iota_X P)$.
We define it as a meromorphic function of $J_X \times W_X^1$ while
Buchstaber, Enolskii, and Leykin defined it as $J_X \times J_X$ \cite{BEL}.
Thus ours is different from the one in \cite{BEL}.

Proposition \ref{5pr:Baker_wp1}, Lemmas \ref{4lm:zetazeta=mero} and \ref{5lm:Baker_wp2} show the generalization of (\ref{4eq:BAeq_g1}).
\begin{proposition}
For $u=\tw(\iota_X(x_1, y_1), \ldots, \iota_X(x_g,y_g))$ and $v=\tw(\iota_X(\tx,\ty))$, we have
\begin{equation}
\left[\frac{\partial^2}{\partial u_g^2}-2\wp_{gg}(u)\right]\psi(u;v)
=(\tx+\lambda_{2g})\psi(u;v),
\label{4eq:BAeq_gg}
\end{equation}
\begin{equation}
\left[\frac{\partial}{\partial u_g}-(\zeta_g(u+v) - \zeta_g(u))\right]\psi(u;v)
=\tzeta_{g}(v) \psi(u;v).
\label{4eq:BAeq_gg'}
\end{equation}
\end{proposition}

\noindent
{\bf{Proof:\ }}
The same as the case of $g=1$.
\qed

\section{The algebraic differential relation of genera two and three}

We extend the algebraic differential relation (\ref{eq:01}) of genus one to genera two and three.

\subsection{Genus two}
Let us consider only the genus $g=2$ case as follows:
We assume $(P_1, P_2, \tP) \in S^2 X\times X$, $(P_i = (x_i, y_i)$, and $\tP=(\tx, \ty)$ such that $u =\tw (\iota_X (P_1, P_2))$ and $v=\tw(\iota_X(\tP))$.

Corollary \ref{cor:deg2} of $g=2$ shows
\begin{equation}
\frac{\sigma(u + v) \sigma(u - v) }
{\sigma(u)^2 \sigma_{\natural_1}(v)^2} =
\tx^2 -\wp_{22} \tx - \wp_{21}.
\label{eq:2g_add}
\end{equation}

We immediately obtain the following relation from (\ref{eq:2g_add}), though we do not use it in this paper, 
\begin{lemma}
Let us consider the case $\tP^{(\pm)}=(\pm\tx, \ty^{(\pm)})$ and $v^{(\pm)}:=w(P^{(\pm)})$.
Then we have
\begin{equation}
\frac{\sigma(u + v^{(+)}) \sigma(u - v^{(+)}) }
{\sigma(u)^2 \sigma_{\natural_1}(v^{(+)})^2} 
-\frac{\sigma(u + v^{(-)}) \sigma(u - v^{(-)}) }
{\sigma(u)^2 \sigma_{\natural_1}(v^{(-)})^2} =-2\wp_{22} \tx.
\label{eq:g2Psi}
\end{equation}
\end{lemma}

Now we state our first main theorem in this paper, in which we obtain the generalization of (\ref{eq:NLSE1}) or (\ref{eq:01}).

\begin{theorem}\label{thm:NLSg2}
For $(P_1, P_2, \tP) \in S^2 X\times X$ such that 1) $x_1 x_2=-\tx^2$, or 2) $x_1 x_2=0$, and $\alpha \in \CC^2$ and $c, \beta \in \CC$, we let
\begin{equation*}
\Psi(u; v, t):=\frac{1}{\sqrt{\tx}}\ee^{\trp\alpha u+\ii\beta t+c}
\psi(u;v),\quad
\Psi^*(u; v, t):=\frac{1}{\sqrt{\tx}}\ee^{-\trp\alpha u-\ii\beta t-c}
\psi(u;-v).
\end{equation*}
The following formulae $(g=2)$ hold for the conditions 1) and 2),
\begin{eqnarray*}
\mbox{1)\ :\ }& &
\left[-\ii\frac{\partial}{\partial t}-\alpha_g \frac{\partial }{\partial u_g}\right]\Psi -\frac{1}{2} 
\frac{\partial^2}{\partial u_g^2}\Psi +(\Psi^*\Psi) \Psi =
\frac12 (2\beta+\alpha_g^2-\tx-\lambda_2)\Psi, \\
\mbox{2)\ :\ }& &
\left[-\ii\frac{\partial}{\partial t}-\alpha_g \frac{\partial }{\partial u_g}\right]\Psi -\frac{1}{2} 
\frac{\partial^2}{\partial u_g^2}\Psi +(\Psi^*\Psi) \Psi =
\frac12 (2\beta+\alpha_g^2-3\tx-\lambda_2)\Psi. \\
\end{eqnarray*}
\end{theorem}

\noindent
{\bf{Proof:\ }}
From (\ref{4eq:BAeq_gg}), we have
$$
\left[-\ii\frac{\partial}{\partial t}-\alpha_g \frac{\partial }{\partial u_g}\right]\Psi -\frac{1}{2} 
\frac{\partial^2}{\partial u_g^2}\Psi +\wp_{gg} \Psi =
\frac12 (2\beta+\alpha_g^2-\tx-\lambda_2)\Psi.
$$
Under the restriction 1), $(\Psi^* \Psi)$ is reduced to $(1/\tx)$ times $\tx \wp_{22}(u)$. whereas under the restriction 2), $(\Psi^* \Psi)$ is equal to $-(\tx-\wp_{22}(u))$.
\qed

\subsection{Genus three}
Let us consider the genus $g=3$ case as follows.
We assume $(P_1, P_2, P_3, \tP) \in S^3 X\times X$, $(P_i = (x_i, y_i)$, and $\tP=(\tx, \ty)$ such that $u =\tw (\iota_X (P_1, P_2, P_3))$ and $v=\tw(\iota_X(\tP))$.

Corollary \ref{cor:deg2} of $g=3$ shows
\begin{equation}
\frac{\sigma(u + v) \sigma(u - v) }
{\sigma(u)^2 \sigma_{\natural_1}(v)^2} =
\tx^3 -\wp_{33} \tx^2 - \wp_{32}\tx - \wp_{31}.
\label{eq:3g_add}
\end{equation}

Now we state our second main theorem in this paper, in which we obtain the generalization of (\ref{eq:NLSE1}) or (\ref{eq:01}).

\begin{theorem}\label{thm:NLSg3}
For $(P_1, P_2, P_3, \tP) \in S^3 X\times X$ such that 
$\tx(x_1x_2+x_2x_3+x_3x_1)-x_1x_2x_3=0$, $\alpha \in \CC^g$ and $c, \beta \in \CC$, we let
\begin{equation*}
\Psi(u; v, t):=\frac{1}{\tx}\ee^{\trp\alpha u+\ii\beta t+c}
\psi(u;v),\quad
\Psi^*(u; v, t):=\frac{1}{\tx}\ee^{-\trp\alpha u-\ii\beta t-c}
\psi(u;-v).
\end{equation*}
The following formula $(g=3)$ holds.
$$
\left[-\ii\frac{\partial}{\partial t}-\alpha_g \frac{\partial }{\partial u_g}\right]\Psi -\frac{1}{2} 
\frac{\partial^2}{\partial u_g^2}\Psi +(\Psi^*\Psi) \Psi =
\frac12 (2\beta+\alpha_g^2-3\tx-\lambda_4)\Psi. \\
$$
\end{theorem}

\noindent
{\bf{Proof:\ }}
From (\ref{4eq:BAeq_gg}), we have
$$
\left[-\ii\frac{\partial}{\partial t}-\alpha_g \frac{\partial }{\partial u_g}\right]\Psi -\frac{1}{2} 
\frac{\partial^2}{\partial u_g^2}\Psi +\wp_{gg} \Psi =
\frac12 (2\beta+\alpha_g^2-\tx-\lambda_4)\Psi. 
$$
Under the restriction, $(\Psi^* \Psi)$ is reduced to $(1/\tx)^2$ times $-(\tx^3-\tx^2 \wp_{33}(u))$.
\qed

\begin{remark}
{\rm{
As mentioned in Introduction the Baker equation (\ref{4eq:BAeq_gg}) contains   the potential $\wp_{gg}$, while the terms $\wp$ in $F(\tx)$ in the addition theorem (\ref{eq:addthem_F}) do not coincide with $\wp_{gg}$ except for $g=1$.
Thus, this discrepancy prevents us from obtaining the generalization of (\ref{eq:01}) directly for a long time.

Even for genera two and three, we found the generalization by restricting the configurations of points in $S^g X$.
In other words, the restriction may be effective for the hyperelliptic solution of the NLS equation in terms of $\sigma$ function, allowing us to easily find a generalization to the general genus.

However we should consider how the restriction effects on the Jacobian $J_X$.
Further we note that in the NLS equation, we have crucial problems related to the reality conditions depending on the {\lq\lq}nonfocusing" and {\lq\lq}focusing" NLS equations \cite{Pr85} or the choice of $(\alpha, u)$ which belongs to $\ii \RR\times \RR$ or $ \RR\times \ii\RR$.
Since these problems are basically complicated as in \cite{Ma23a}, we will consider them as the next steps. 
}}
\end{remark}

\subsection*{Acknowledgments:}
The author is grateful to Professor Emma Previato for suggestions on this study at a very early stage and to Dr. Kayo Kinjo for discussing the algebraic solution of the NLS equation based on \cite{BBEIM}.
The author has been supported by the Grant-in-Aid for Scientific Research (C) of the Japan Society for the Promotion of Science Grant, No.21K03289.

\bigskip

\end{document}